\documentclass[prb,twocolumn,showpacs,preprintnumbers,amsmath,amssymb]{revtex4}
\usepackage{graphicx}
\usepackage{dcolumn}
\usepackage{epsfig}
\usepackage{bm}

\begin{document}

\title{Low energy states with different symmetries in
 the $t$-$J$ model with two holes
on a 32-site lattice}
\author{P. W. Leung}
\email{P.W.Leung@ust.hk}
\affiliation{Physics Dept., Hong Kong University of Science and Technology,
Clear Water Bay, Hong Kong}
\date{\today}

\begin{abstract}
We study the low energy states of the $t$-$J$ model with two holes on
a 32-site lattice with periodic boundary conditions. In contrary to
common belief, we find that the
state with $d_{x^2-y^2}$ symmetry is not always the ground state
in the realistic  parameter range  $0.2\le J/t\le 0.4$. 
There exist low-lying finite-momentum
$p$-states whose energies are lower than the $d_{x^2-y^2}$ state when
$J/t$ is small enough. We
compare various properties of these low energy states at
$J/t=0.3$ where they are almost degenerate, and 
find that those properties associated with the holes
(such as the hole-hole correlation and the
electron momentum distribution function) are very different between
the $d_{x^2-y^2}$ and $p$ states, while their spin
properties are very similar.
Finally, we demonstrate that by adding ``realistic'' terms to the
$t$-$J$ model Hamiltonian, we can easily destroy the $d_{x^2-y^2}$
ground state. This casts doubt on the robustness of the $d_{x^2-y^2}$
state as the ground state in a microscopic model for the high
temperature superconductors.
\end{abstract}

\maketitle

\section{Introduction}

The $t$-$J$ model was proposed as a microscopic model to describe the
low energy physics of high temperature superconductors. It mimics
the doped CuO$_2$ planes found in high temperature superconductors by a
system consisting of mobile holes moving in a spin background on a
two-dimensional square lattice.\cite{zr88} One possible mechanism for
superconductivity is the Bose-Einstein condensation of  ``preformed'' hole
pairs. To see whether this is
possible, first we have to understand how the
holes interact. It is well-known that a mobile hole causes frustration
in the antiferromagnetic spin background. Longer range interaction
between the holes must be mediated through the spin distortion they produce.
Various studies\cite{d94}
 seem to indicate that in the ``realistic'' parameter range
$J/t\sim 0.3$--0.4, the holes form bound pairs
in the $d_{x^2-y^2}$ channel. Theoretically this  can be
understood as  resulting from the magnon exchange interaction which selects the
$d_{x^2-y^2}$ channel over others.\cite{magnon}
Alternatively,  the density
matrix renormalization group (DMRG) approach has been used to 
work out the spin structure in the
vicinity of a hole pair in real space.\cite{ws97} 
It was found that by pairing up
at a distance of $\sqrt{2}$, the two holes can share the spin
frustration and is therefore energetically favorable. 
This is consistent with the numerical
observation\cite{horsch,p94,I}  that the holes have the
largest probability of being at  $\sqrt{2}$ apart. Later it was
shown  that such pairing of holes at $\sqrt{2}$  arises
naturally if the wavefunction has $d_{x^2-y^2}$ symmetry.\cite{rd98}
But in order to firmly establish the role of $d$-wave hole pairing
in the theory of superconductivity, one has to understand how
strong and robust the pairing is. In fact numerical study\cite{I} has shown
that the binding energy of the $d$ channel is negative but small, and
that the holes may not be tightly bound in real space.
Recently the question on the robustness of $d$-wave pairing
 was addressed by using the anisotropic $t$-$J_z$
model as the starting point.\cite{cl99} It concluded that
different mechanisms may select states with different symmetries
 and their competition
may destroy the $d$-wave ground state. 
This raises doubt on the robustness of the $d$-wave ground state
because it is known that the $t$-$J$ model alone is not enough to
explain fully the hole dynamics of Sr$_2$CuO$_2$Cl$_2$ measured
 by angle-resolved
photoemission.\cite{wells} In order to reproduce the measured spectral
functions, one has to add longer-range hole hopping terms ($t'$,
 $t''$ and a three-site hopping term) 
to the $t$-$J$ model Hamiltonian.\cite{bcs96,nvhdg95,lwg97}

Motivated by the above discussion, we study the low energy states of
the two-hole $t$-$J$ model with
different symmetries using exact diagonalization
(ED). Previous ED studies were mostly performed on square lattices of
16, 20, and 26 sites.
Since we are focusing on the symmetry of the wavefunctions, the
geometry of the lattice is very important. For example, it is
well-known that on the 16-site ($4\times4$) lattice the system has
spurious degeneracy at the wavevectors $(\pi/2,\pi/2)$ and $(\pi,0)$. 
The 32-site lattice is free from these
spurious effects. 
The $d_{x^2-y^2}$ state of the two-hole model on the
32-site lattice has been studied in Ref.~\onlinecite{I}. Besides,
previous works have indicated that there is a $p$-wave state whose
energy is very close to that of the $d_{x^2-y^2}$ state at realistic
values of $J/t$.\cite{l00}
The existence of this low energy state
casts doubt on the robustness of the $d_{x^2-y^2}$ state as the ground
state of the two-hole model. The energy difference is so small
that any additional terms in the Hamiltonian may change the ground
state from one to another. 
This cross-over may be realized by adding  longer range hopping
terms mention above
to the $t$-$J$ model Hamiltonian. Furthermore, it has been shown that
such terms favor the $p$-wave state.\cite{cl99} This shows that the
physics of the low-lying $p$ state may not be irrelevant.
Therefore it is important to understand
the similarities and differences between the properties of the
$d_{x^2-y^2}$ and the $p$
states. In a previous publication\cite{l00} we have shown that the
hole and current correlations of these two states are very different.
In this paper, we will discuss in more detail the low-lying states of
the two-hole model in a larger range of $J/t$. The properties of these
states will be compared at the realistic value of $J/t=0.3$. Finally,
we will demonstrate the effects of longer-range hopping and 
short-range Coulomb repulsion on the symmetry of the ground state.

\section{Low-lying states with different symmetries}
\label{sec:energy}

The Hamiltonian of the $t$-$J$ model is
\begin{equation}
{\cal H} = -t\sum_{\langle ij\rangle\sigma}(\tilde{c}^\dagger_{i\sigma}
\tilde{c}_{j\sigma}+{\rm H.c.})+J\sum_{\langle ij\rangle} 
({\bf S}_i\cdot {\bf S}_j
-\frac{1}{4}n_in_j),
\label{hamiltonian}
\end{equation}
where $\tilde{c}^\dagger$ and $\tilde{c}$ are the projected fermion
operators, and $n_i\equiv\tilde{c}_i^\dagger\tilde{c}_i$ is the fermion
number operator.
We solve this model with two holes on a 32-site lattice
using the standard Lanczos algorithm.
To find the low-energy states, we consider subspaces
with different rotational symmetries and momenta. We concentrate on
the rotational symmetries $s$, $p$, and $d$. States having
larger angular momentum usually have higher energy and
are not relevant in our study. The momenta of the subspaces we study
are $(0,0)$, $(\pi,\pi)$ and $(\pi,0)$. The dimensions of the 
subspaces with $s$ and
$d$  symmetries are about 150 million, and that of the subspaces with
$p$ symmetry 
are about 300 million. Most of the calculations are performed on a
cluster of commodity personal computers. In  subspaces with $d$
symmetry, we
achieve a speed of about 4.3 minutes per Lanczos iteration on a
32-node cluster, where each node utilizes a 1 GHz Althon CPU.

\begin{figure}
\centerline{
\psfig{figure=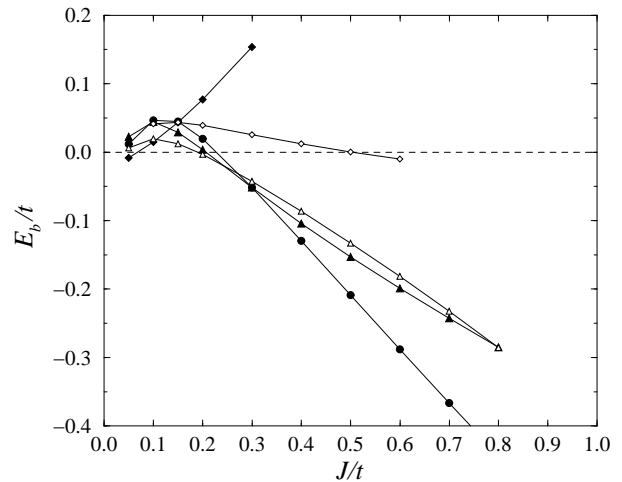,width=8cm}
}
\caption{Two-hole binding energies of the $s$ (solid diamonds), 
$d$ (solid circles),
$p_{(\pi,\pi)}$ (solid triangles), $p_{(\pi,0)}$ (empty triangles),
and $p_{(0,0)}$ (empty diamonds) states as defined
in the text.}
\label{fig:delta2}
\end{figure}

The energies of the low-lying states are tabulated in Table~\ref{tab:gsenergy}.
\begin{table*}
\caption{Energies ($E_{2h}/t$)  of the low-lying states of the
  two-hole $t$-$J$ model on a 32-site lattice at different $J/t$.
The $s$, $d$, $p_{(0,0)}$, $p_{(\pi,\pi)}$ and $p_{(\pi,0)}$ states 
are defined in the text. Also
shown are the ground state energies $E_{1h}/t$ of the one-hole model. The
ground state energy of the undoped model is $E_{0h}=-37.7657342J$.}
\label{tab:gsenergy}
\begin{ruledtabular}
\begin{tabular}{ddddddd}
\multicolumn{1}{c}{$J/t$}& \multicolumn{1}{c}{$E_{1h}/t$} & \multicolumn{5}{c}{$E_{2h}/t$} \\
&&\multicolumn{1}{c}{$s$} & \multicolumn{1}{c}{$d$} & \multicolumn{1}{c}{$p_{(0,0)}$} & \multicolumn{1}{c}{$p_{(\pi,\pi)}$} & \multicolumn{1}{c}{$p_{(\pi,0)}$} \\
\colrule
0.1&  -6.419382     &    -9.047472  &-9.015703 &   -9.0208259142 &        -9.017720   &    -9.042840  \\
0.2&  -9.766819     &    -11.903649 &-11.960871&   -11.9412605733&        -11.976890  &    -11.983860 \\
0.3& -13.161933     &    -14.840663 &-15.045603&   -14.9685556227&        -15.045602  &    -15.036724 \\
0.4& -16.583058     &               &-18.189385&   -18.0477851139&        -18.163935  &    -18.146186 \\
0.5& -20.021272     &               &-21.368517&   -21.1594455768&        -21.312775  &    -21.292519 \\
0.6& -23.471807     &               &-24.572086&   -24.2941123207&        -24.483247  &    -24.465752 \\
0.7& -26.931796     &               &-27.793984&                 &        -27.670397  &    -27.660030 \\
0.8& -30.399375     &               &-31.030391&                 &        -30.871142  &    -30.871578 \\
\end{tabular} 
\end{ruledtabular}
\end{table*}
We are interested in the two-hole binding energy which is defined as
\begin{equation}
E_b=E_{2h}-2E_{1h}+E_{0h},
\end{equation}
where $E_{nh}$ is the energy of the $n$-hole system.
The ground state of the undoped system with energy $E_{0h}$ is a
totally symmetric state with momentum $(0,0)$.
The ground state of the one-hole system with energy $E_{1h}$ has
momentum $(\pi/2,\pi/2)$.
$E_b$ is the energy gain of the two-hole system relative to two
one-hole systems, and therefore indicates
the relative tendency for the two holes to form a bound pair.
Keeping in mind that the ``realistic'' value
of $J$ is roughly between $0.2t$ and $0.4t$,
we calculate the binding energies in a larger range
of $J$ in order to get a clear picture of the
crossing-over of the energy levels.
Fig.~\ref{fig:delta2} shows $E_b$ of the low-lying states with $s$,
$p$ and $d$ symmetries. 
The lowest energy state in the subspace with $s$ symmetry is a spin singlet
with momentum $(0,0)$. We call this the $s$ state. 
At small $J$ ($J\leq 0.1t$), this is the ground
state. But its energy rises very fast with $J$ and
soon changes from the ground state to an excited state. This shows
that the $s$ channel is energetically unfavorable and is irrelevant
at the realistic parameter $J\sim0.3t$.
The lowest energy state with $d$ symmetry is a spin singlet with
momentum $(0,0)$ and $d_{x^2-y^2}$ symmetry. We call this the $d$ state.
This is the same state reported in Ref.~\onlinecite{I}. At $J>0.3t$, it
is the ground state of the two hole system. Note that the binding
energy of this state scales
linearly with $J$ when $J>0.3t$, 
reflecting the magnetic nature of the hole binding
mechanism.\cite{dry90}
Those states with $p$ symmetry are more complicated. In
Fig.~\ref{fig:delta2} we show two low energy states with $p$ symmetry. Both
of them are two-fold degenerate and are spin singlets. One has
momentum $(\pi,\pi)$. We call it the $p_{(\pi,\pi)}$ state. The other has
momentum $(\pi,0)$.
We call it the $p_{(\pi,0)}$ state.\cite{pstate} 
In addition, we also show a higher energy $p$ state with zero
momentum, which we call the $p_{(0,0)}$ state. This state was
mentioned in a previous study using a smaller lattice.\cite{dry90}
Also note that it is a triplet state.
Fig.~\ref{fig:delta2} indicates a general trend that as $J$ decreases,
the energy of the $d$ state is pushed up faster than the $p$ states.
While at $J>0.3t$ the $p_{(\pi,\pi)}$ and $p_{(\pi,0)}$ states
are clearly excited states, the energy levels look quite
complicated in the range $0.2t\le J\le 0.3t$. Note that the $d$, $p_{(\pi,\pi)}$
and $p_{(\pi,0)}$ states are almost degenerate at $J=0.3t$, although the $d$
state is still the ground state. When $J$ is slightly smaller than $0.3t$,
it looks like the $p_{(\pi,\pi)}$ state has lower energy than the $d$ state but
at $J=0.2t$, the $p_{(\pi,0)}$ state has the lowest energy among the three.
At smaller $J$, we anticipate more serious finite-size effects.
Based on our data only we cannot conclude where the
crossing-overs of the energy levels occur, nor 
whether they exist in the thermodynamic limit.
But the fact that the crossing-overs in our system occur at $J$ values
within  the realistic range is sufficient to cast doubt on the
symmetry of the ground state in the two-hole model.
Regardless of whether one of the $p$ states actually becomes the
ground state at certain range of $J$, 
their existence as low-lying states
is unquestionable. Furthermore, since the energy levels of these
states are very close, inclusion of farther-than-nearest-neighbor
hopping terms or Coulomb repulsion may easily change the ground state from
one to another. Therefore we think that they are relevant and warrant
detail studies.
In the following sections we study in detail the properties of these
states at $J=0.3t$ where they are almost degenerate.

\section{Real space structure of the hole pair}

\begin{figure}
\centerline{
\psfig{figure=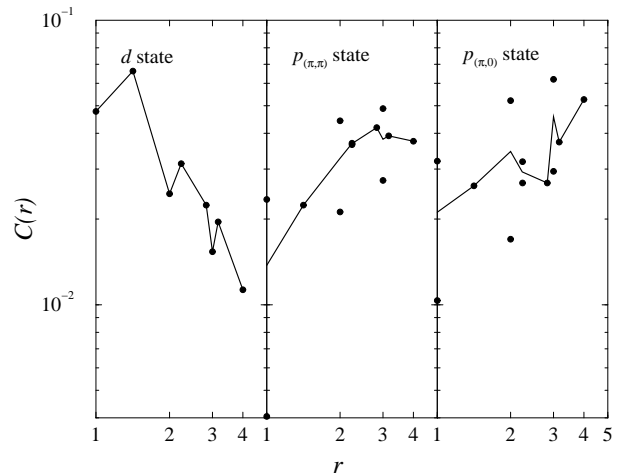,width=8cm}
}
\caption{Hole-hole correlation function $C(r)$ of the
$d$ and $p$ states at $J=0.3t$. Due to the lack of the
rotational symmetry $C_4$ (i.e., a rotation by $\pi/2$) in the $p$
states, there may be two inequivalent points at the same $r$.  In this
case their
values are indicated by the symbols and the solid line joins their mean values.
}
\label{fig:hh_corr}
\end{figure}

Despite the fact that the binding energies of the $d$ and $p$
states at $J=0.3t$ are very similar and negative,
their hole-hole correlations are very different.
In Fig.~\ref{fig:hh_corr} we plot the hole-hole correlation function
$C(r)\equiv\langle(1-n_0)(1-n_r)\rangle$.
As already discussed in Ref.~\onlinecite{I}
 and also shown in Fig.~\ref{fig:hh_corr},
the holes in the $d$ state attract each other, with $C(r)$
having a maximum at $r=\sqrt{2}$. This has been observed in previous
ED studies using smaller lattices.\cite{horsch,p94}
Note that the observed attraction cannot be considered to be strong. The
probability $P(r)$ of the holes being at $\sqrt{2}$
[$P(\sqrt{2})=0.26476$]  and
$\sqrt{5}$ [$P(\sqrt{5})=0.25037$] are almost
the same.\cite{prob}
The root-mean-square separation of
the hole pair is $r_{\rm rms}\equiv\sqrt{\langle r^2\rangle}=2.05865$.
Compared to the $r_{\rm rms}$ of two uncorrelated holes in this
lattice, which is $2.38273$, the hole pair in the $d$ state is not
tightly bound. This is consistent with its barely negative $E_b$.
On the contrary the holes in both $p$ states seem to be mutually
repulsive despite the negative $E_b$. Their $r_{\rm rms}$ are
$2.53755$ and $2.53500$ respectively, which are significantly larger
than that of the $d$ state. This shows that the hole pairs in the
$p_{(\pi,\pi)}$ and $p_{(\pi,0)}$
states are unbound. 

\begin{figure}
\centerline{
\psfig{figure=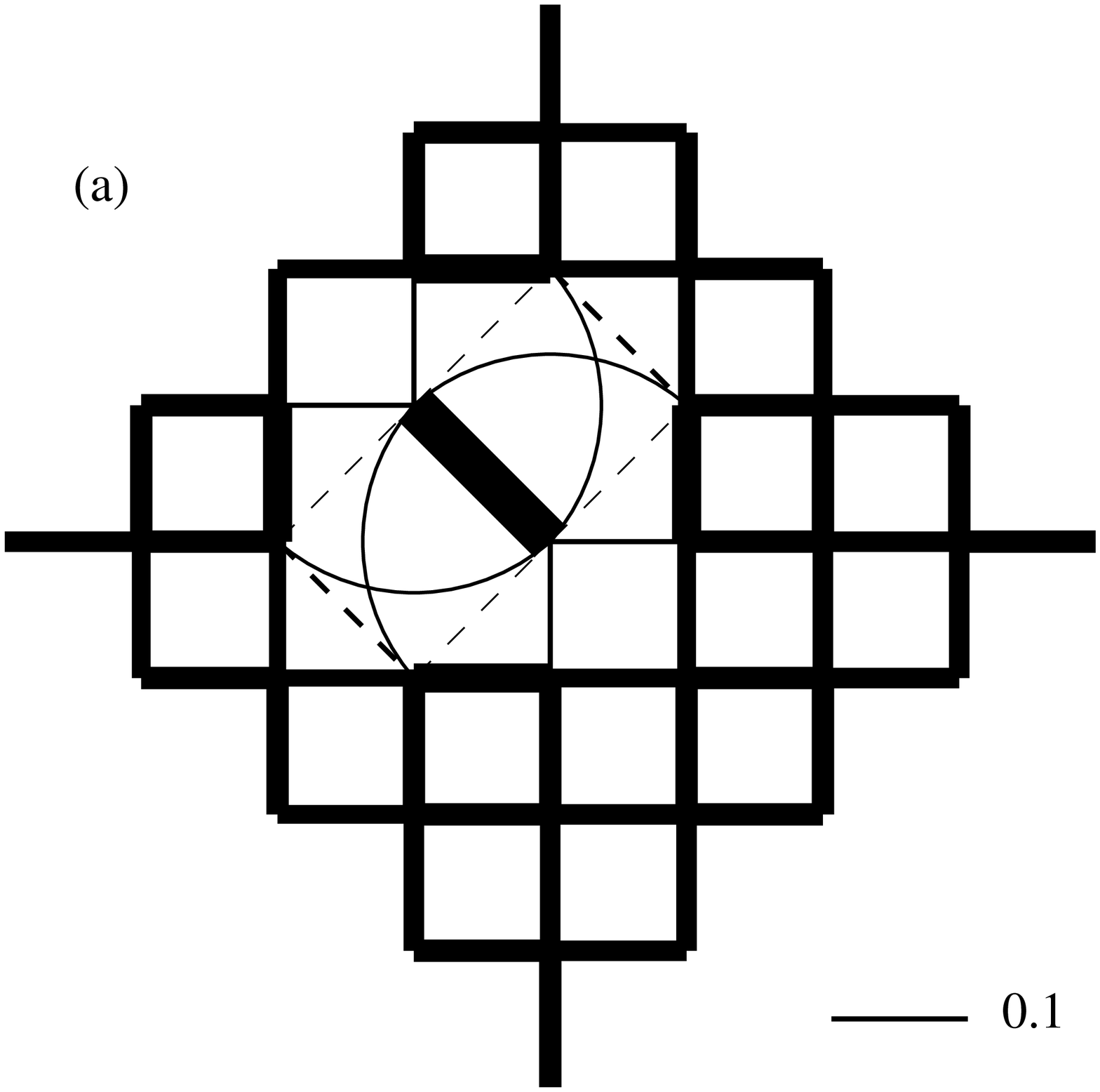,width=4cm}
\psfig{figure=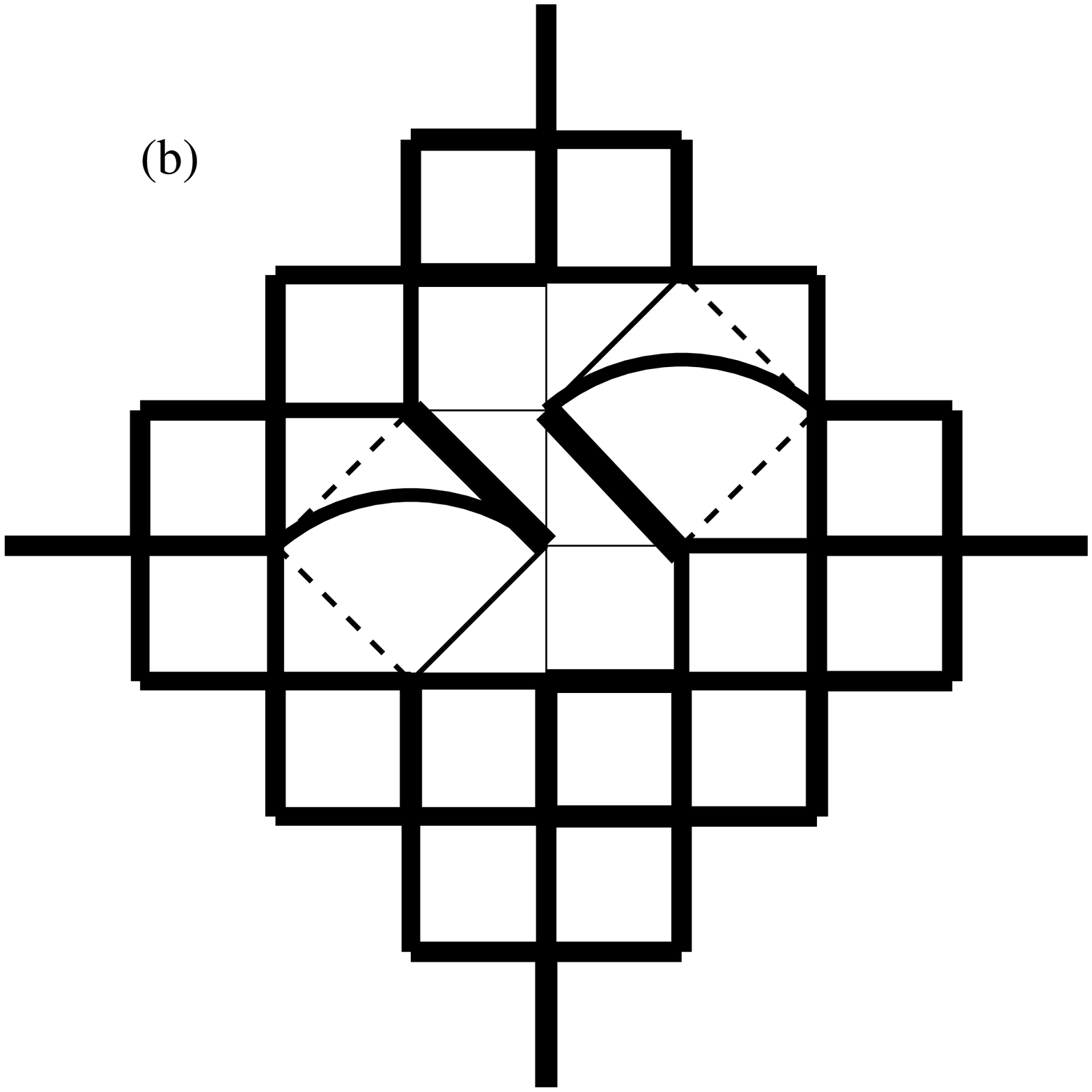,width=4cm}
}
\centerline{
\psfig{figure=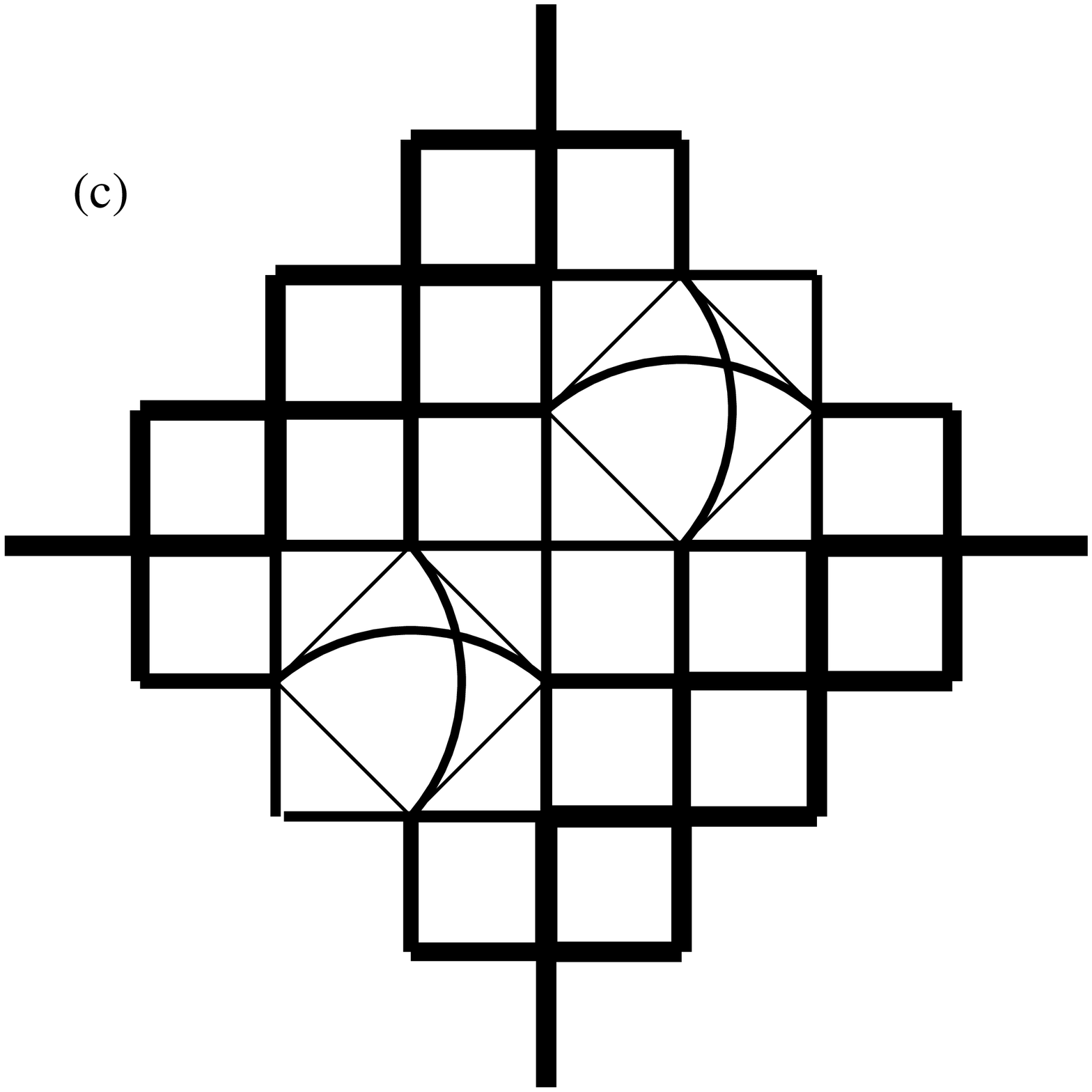,width=4cm}
\psfig{figure=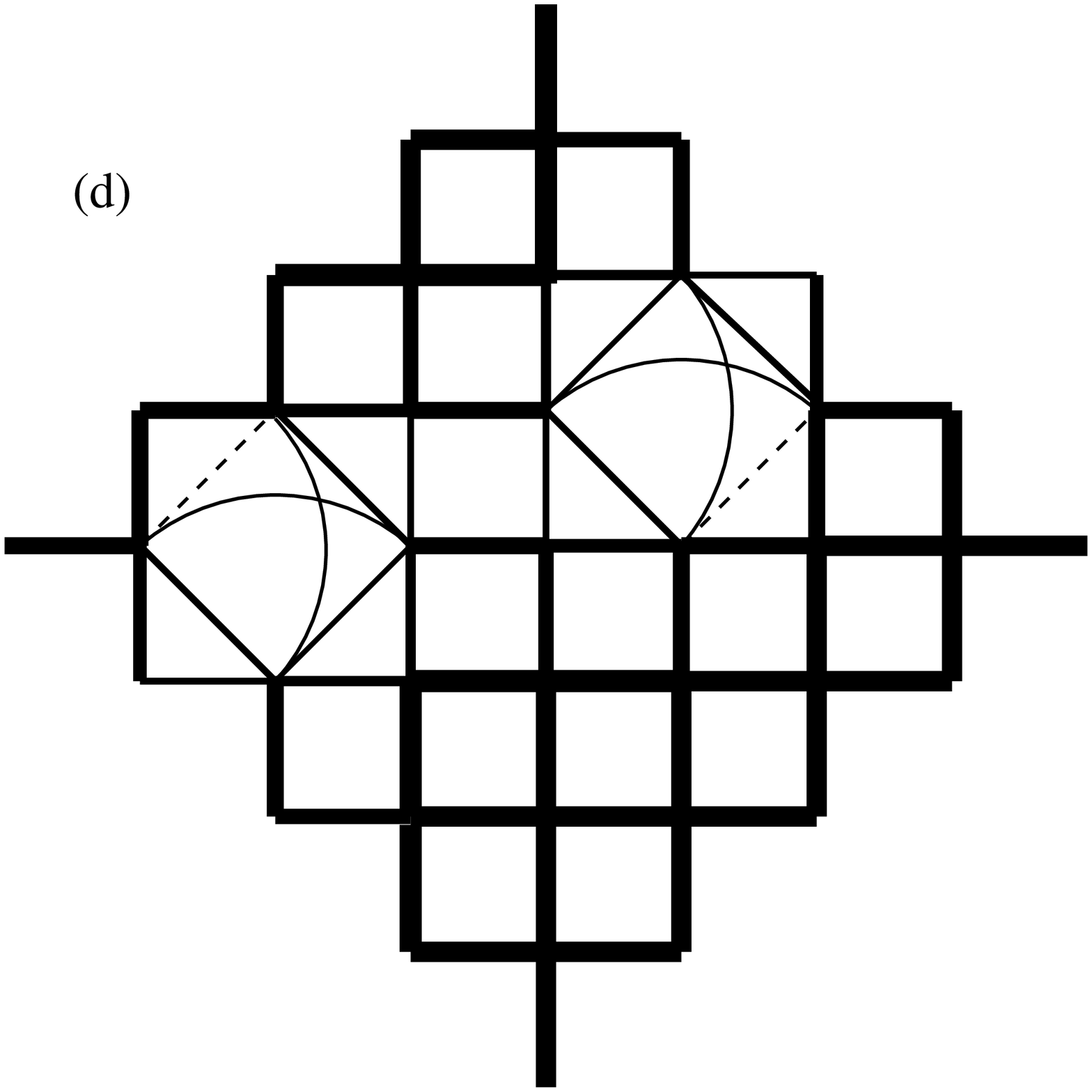,width=4cm}
}
\caption{Spin correlation $\langle{\bf S}_i\cdot{\bf S}_j\rangle$ in
  the $d$ state at $J=0.3t$. The two empty lattice points in each
  diagram are the locations of the holes. 
Only  nearest neighbor spin correlations are shown
  except in the vicinity of the holes where the correlations between spins
  at  $\sqrt{2}$ and 2 apart are also shown.
  The width of the line joining
  two spins $i$ and $j$ is proportional to $\langle{\bf S}_i\cdot{\bf
  S}_j\rangle/P(r)$, where $P(r)$ is the probability of finding the
  holes at the relative locations as shown. Solid and broken lines
  mean $\langle{\bf S}_i\cdot{\bf S}_j\rangle$ are negative and
  positive respectively.}
\label{fig:sisj}
\end{figure}

\begin{figure}
\centerline{
\psfig{figure=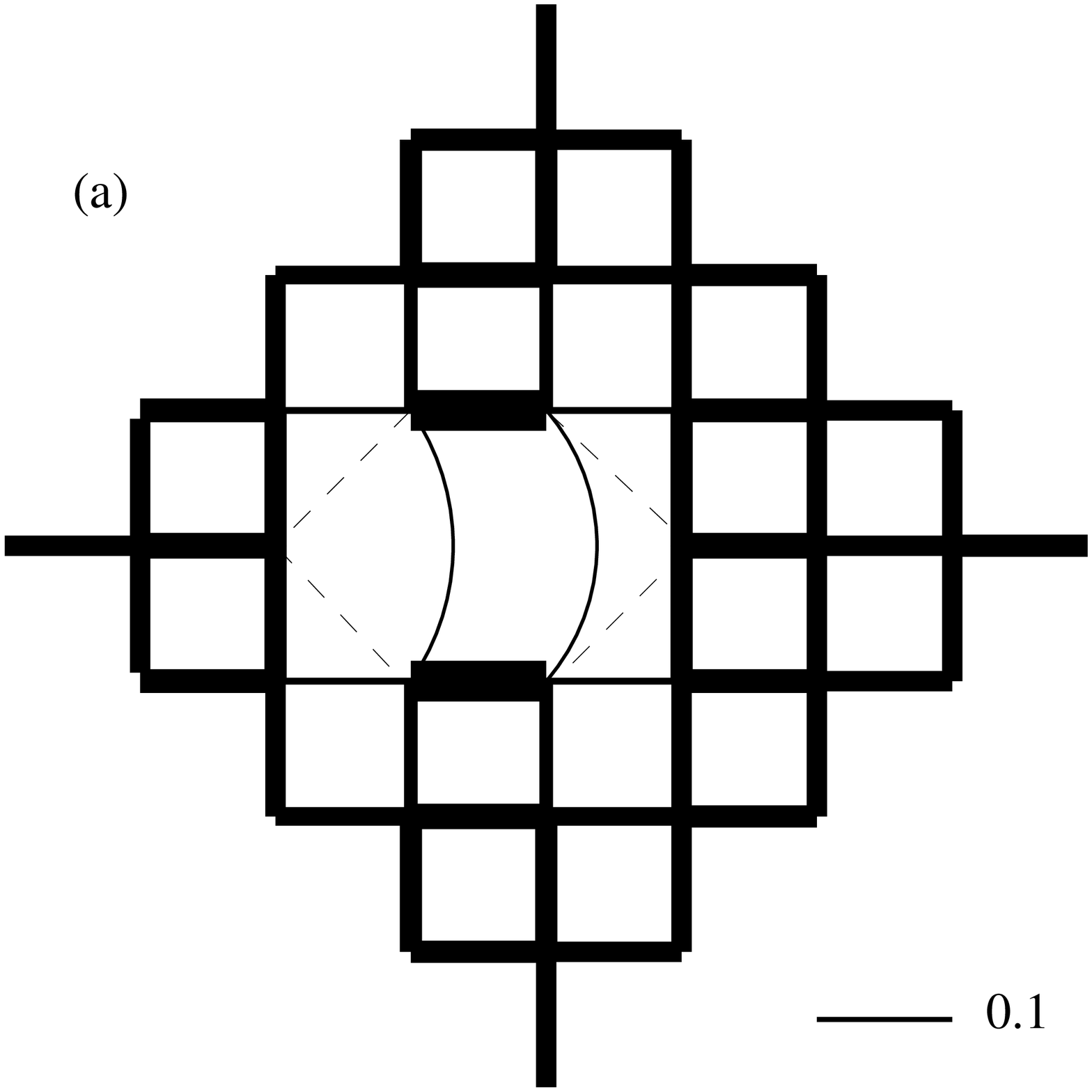,width=4cm}
\psfig{figure=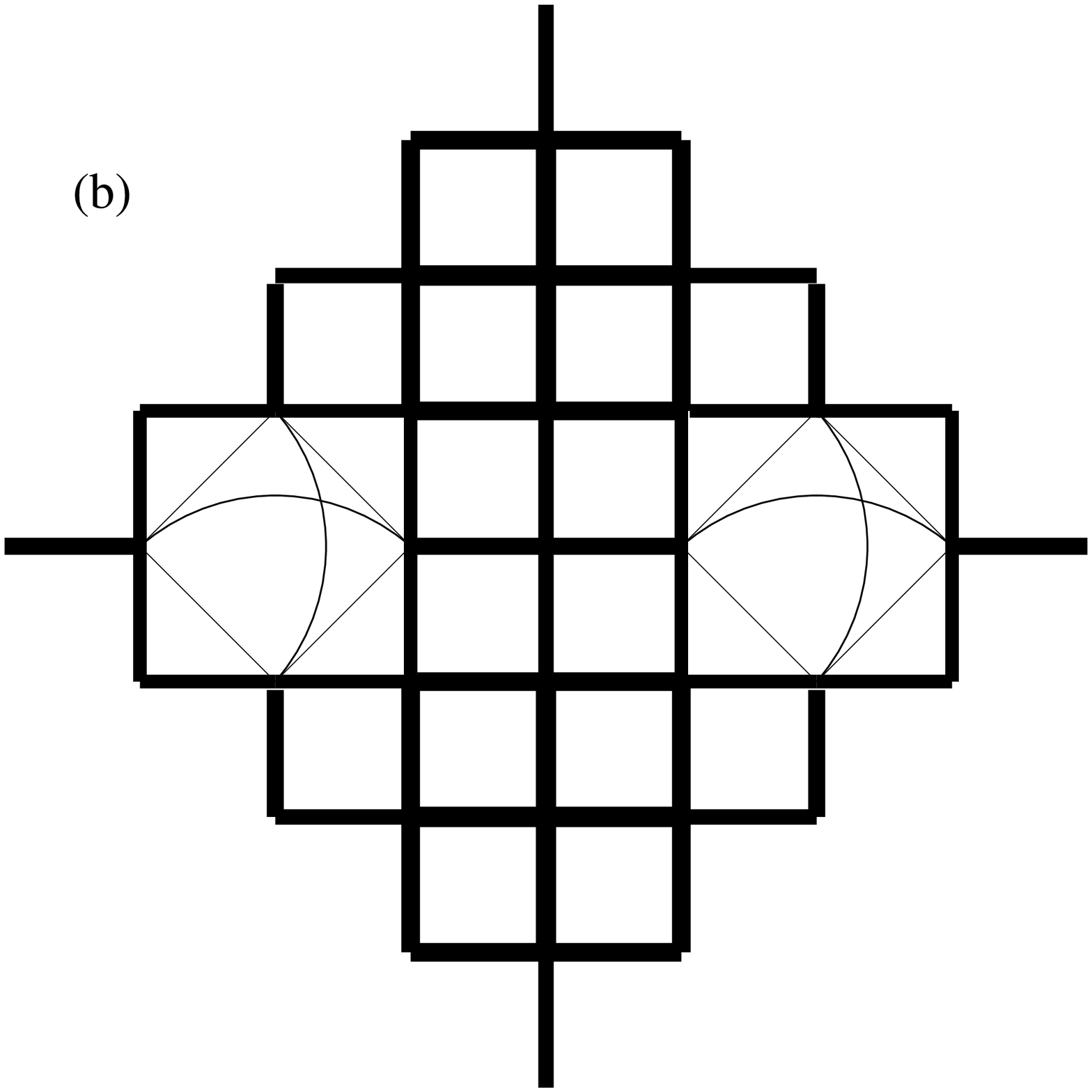,width=4cm}
}
\caption{Same as Fig.~\ref{fig:sisj} but for different hole
  configurations.}
\label{fig:sisj1}
\end{figure}

\begin{figure}
\centerline{
\psfig{figure=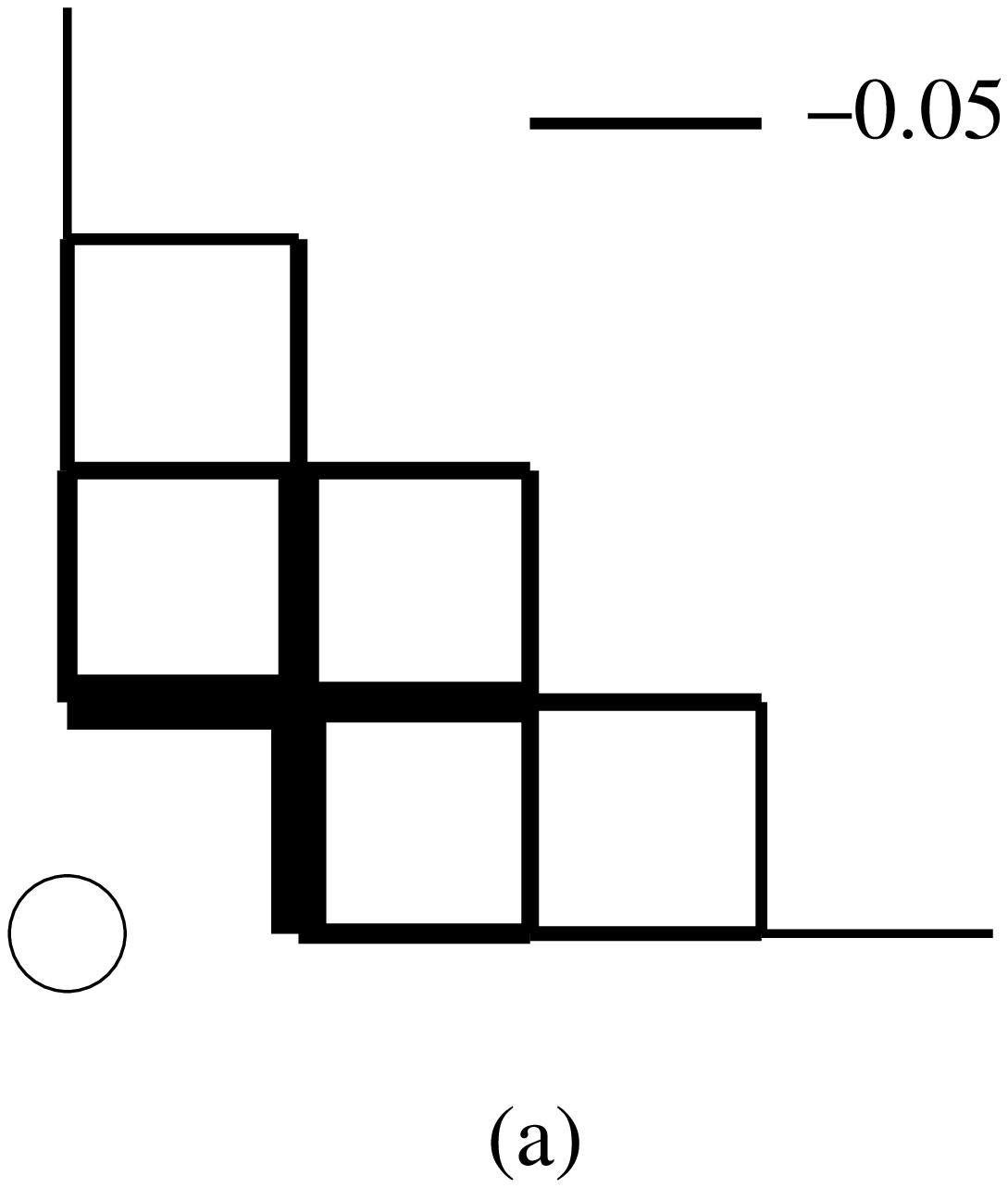,width=3cm}\hspace{1cm}
\psfig{figure=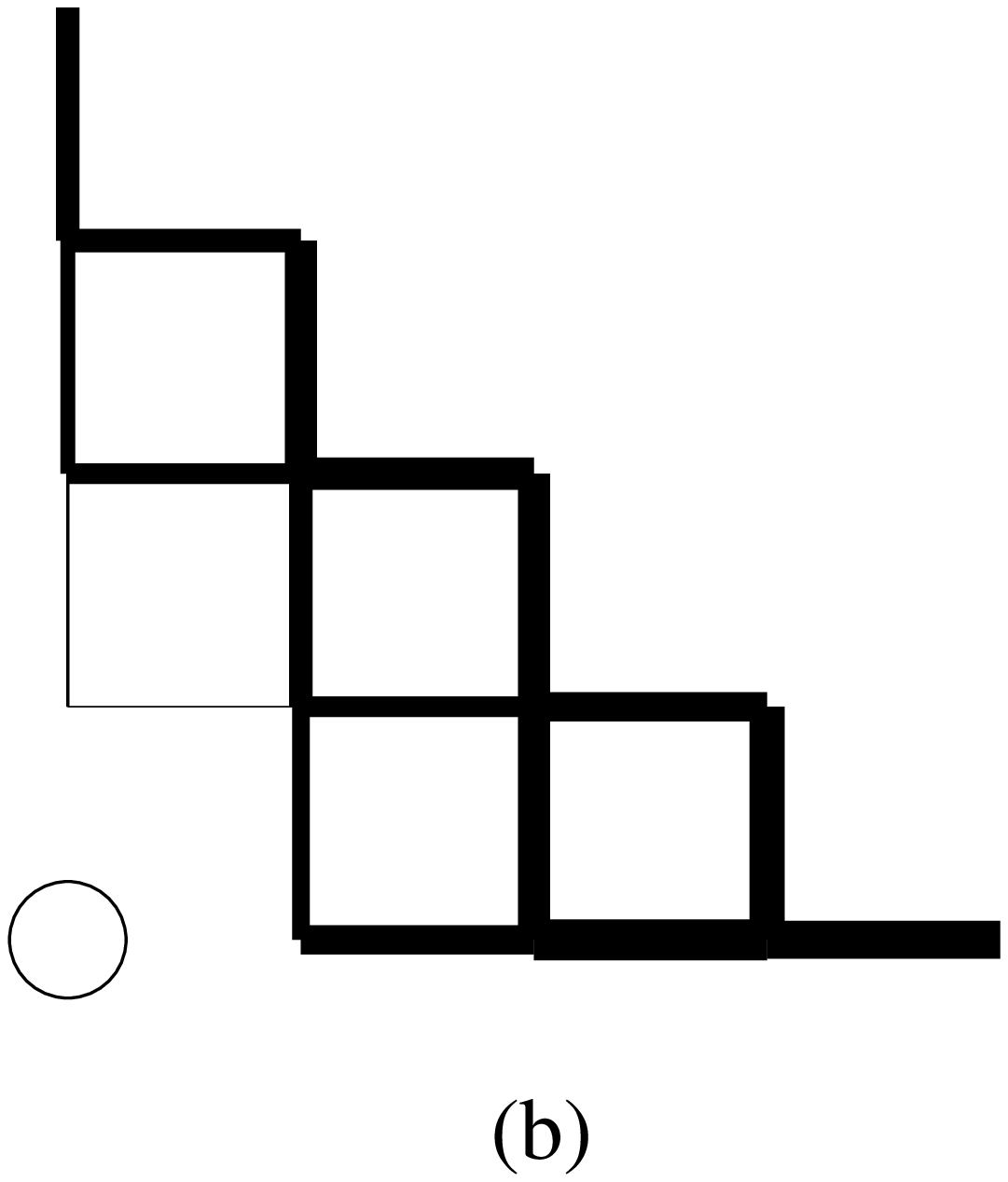,width=3cm}
}
\caption{The hole hopping energy $\langle
  \tilde{c}^\dagger_i\tilde{c}_j\rangle$  when the other hole is fixed at the
  location indicated by an empty circle, in (a) the $d$ state and (b) 
  the $p_{(\pi,\pi)}$ state,
  both at $J=0.3t$.  Only a quarter of the lattice is shown. Values
  at other nearest neighbor pairs $\langle ij\rangle$ can be found by using the
  symmetry of the lattice. }
\label{fig:ke}
\end{figure}

\begin{figure}
\centerline{
\psfig{figure=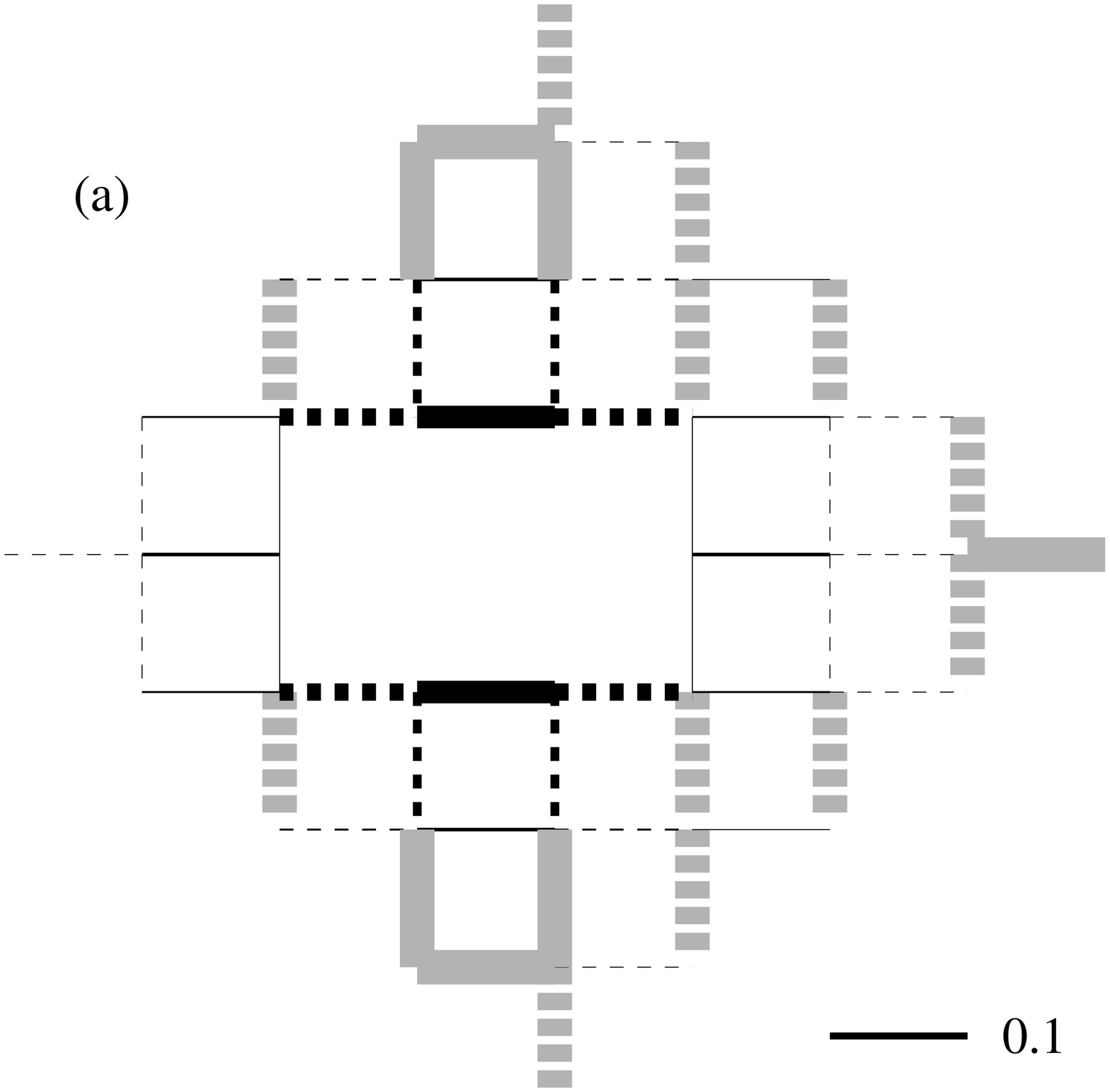,width=4cm}
\psfig{figure=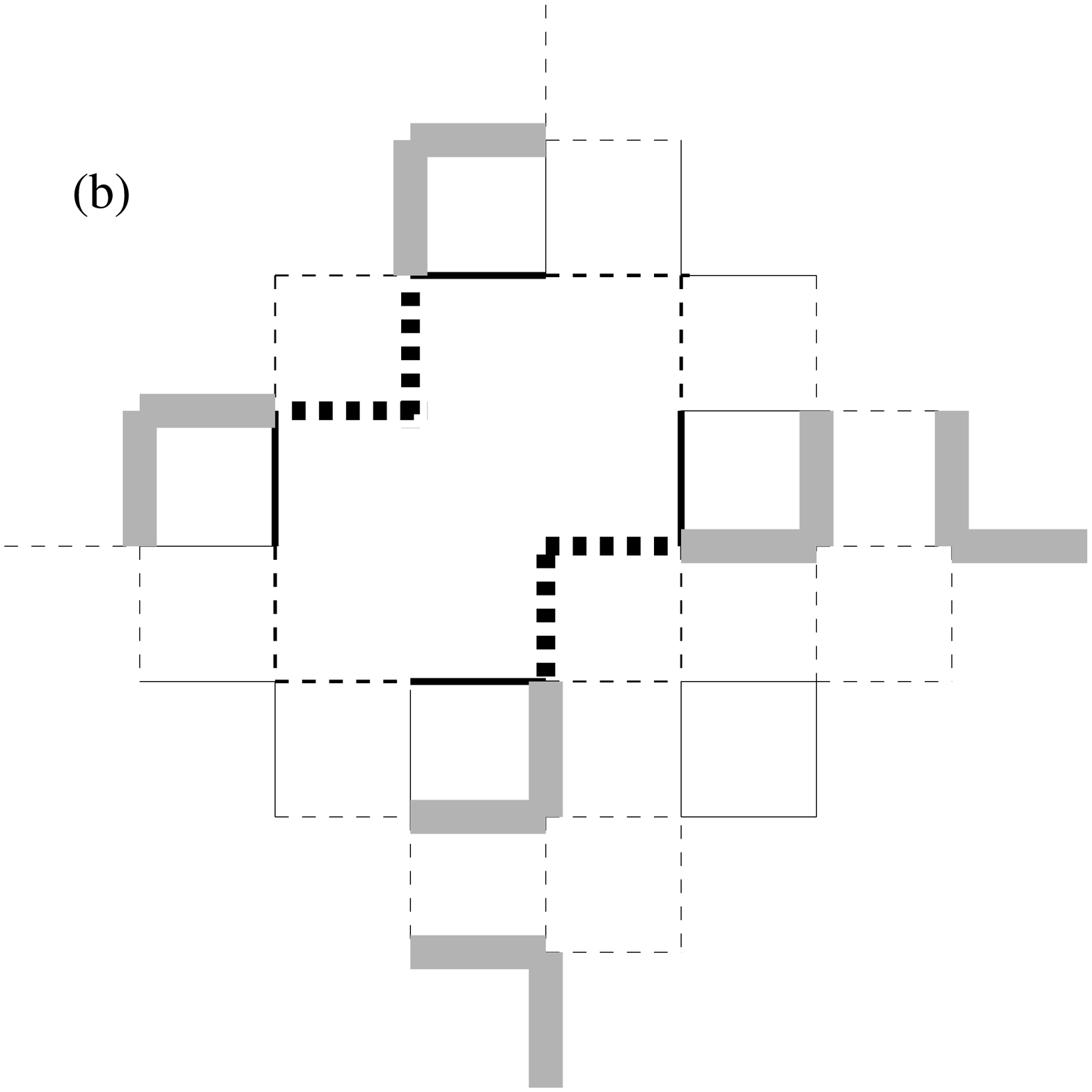,width=4cm}
}
\caption{The spin correlation relative to the undoped value, $\langle{\bf S}_i\cdot{\bf S}_j\rangle/P(r)-\langle{\bf
  S}_i\cdot{\bf S}_j\rangle_{\rm undoped}$, of the $d$ state at
  $J=0.3t$. $\langle{\bf
  S}_i\cdot{\bf S}_j\rangle_{\rm undoped}=-0.34009$. Solid and broken lines
  mean that the singlet correlation are enhanced (more negative) and
  suppressed (less negative) respectively
  relative to $\langle{\bf
  S}_i\cdot{\bf S}_j\rangle_{\rm undoped}$. Only nearest neighbor
  correlations are shown. Those thick shaded lines
  mean that their values are too small to be shown in this scale, with
  magnitude smaller than $0.02$. Their thickness do not represent the
  magnitude. They are drawn to show the sign of their values only.}
\label{fig:dimer}
\end{figure}

The spatial correlation of the
hole pairs described above can be understood in terms of the
nearest neighbor spin correlation $\langle {\bf S}_i\cdot{\bf
  S}_j\rangle$ at fixed hole configuration. 
This analysis was first carried out on a two-hole bound state
using the technique
of density matrix renormalization group,\cite{ws97} and later
by ED.\cite{rd98} Fig.~\ref{fig:sisj} shows our result for the $d$
state on a 32-site lattice.
The most intriguing feature is the existence of a
strong next-nearest neighbor (i.e. across the diagonal of a square
plaquette) singlet bond in between two holes when they are at
distance $\sqrt{2}$ apart. This  is shown in
Fig.~\ref{fig:sisj}a.
It has been argued\cite{ws97}
that such singlet bond increases the hopping overlap between different
hole configurations -- when a hole hops this singlet bond becomes a strong
nearest neighbor bond, and therefore is energetically favorable. But
such singlet bond between two spins in the same sublattice
causes spin frustration around the plaquette.
Therefore the two holes tend to stay at next-nearest-neighbor distance
in order to share the frustrating singlet bond. This explains why the
holes are more likely to be found at distance $\sqrt{2}$ as shown in
Fig.~\ref{fig:hh_corr}.  
When the holes are farther part as shown in Fig.~\ref{fig:sisj}b,
a diagonal singlet bond still exists in the immediate vicinity of each
hole, although it is weakened compared to the one in Fig.~\ref{fig:sisj}a.
Besides, across each hole there is another
strong singlet bond between two spins at a distance of 2.
These two singlet bonds encourage the holes to hop towards each other
because doing so will create two strong nearest-neighbor singlet bonds.
Such characteristic is retained in the other hole configurations shown
in Fig.~\ref{fig:sisj}, although the strength of the singlet bonds
weaken  quickly as the holes
move farther apart.
Note that this set of hole
configurations consistently have larger hole correlation 
$C(r)$ as shown in
Fig.~\ref{fig:hh_corr}. For the other hole configurations where the
holes are along the $x$ or $y$ directions, the characteristics of the
spin correlations around the holes are very different. Two of these
hole configurations are shown in Fig.~\ref{fig:sisj1}. When the holes
are at nearest neighbor distance (Fig.~\ref{fig:sisj1}a), there are
two strong nearest neighbor singlet bonds immediate above and below
the hole pair. 
When one hole hops to become the hole
configuration in Fig.~\ref{fig:sisj}a, one of these singlet bonds
 becomes the strong diagonal single bonds, and the other remains as a
 strong nearest neighbor bond. We can see the reminiscence of this
 bond on the four sides of the diagonal bond in Fig.~\ref{fig:sisj}a.
Such hopping is strongly
preferred as shown in the kinetic energy plot of Fig.~\ref{fig:ke}.
When the holes are at $\sqrt{2}$ apart, the hopping energy
along the four directions are significantly larger.
When the holes are far apart (Fig.~\ref{fig:sisj1}b), the spin
correlation around the hole does not seem to possess any special
feature. The hole hopping energy also decreases rapidly.
However, contrary to a previous study,\cite{ws97}
 when the holes are close to each other, we do not see 
evidence for dimerization in the vicinity of the holes. In
Fig.~\ref{fig:dimer} we show the spin correlation relative to the
undoped value. Except at the immediate
vicinity of the holes, the spin correlation is rather unaffected and
there is no clear evidence of dimerization.

\begin{figure}
\centerline{
\psfig{figure=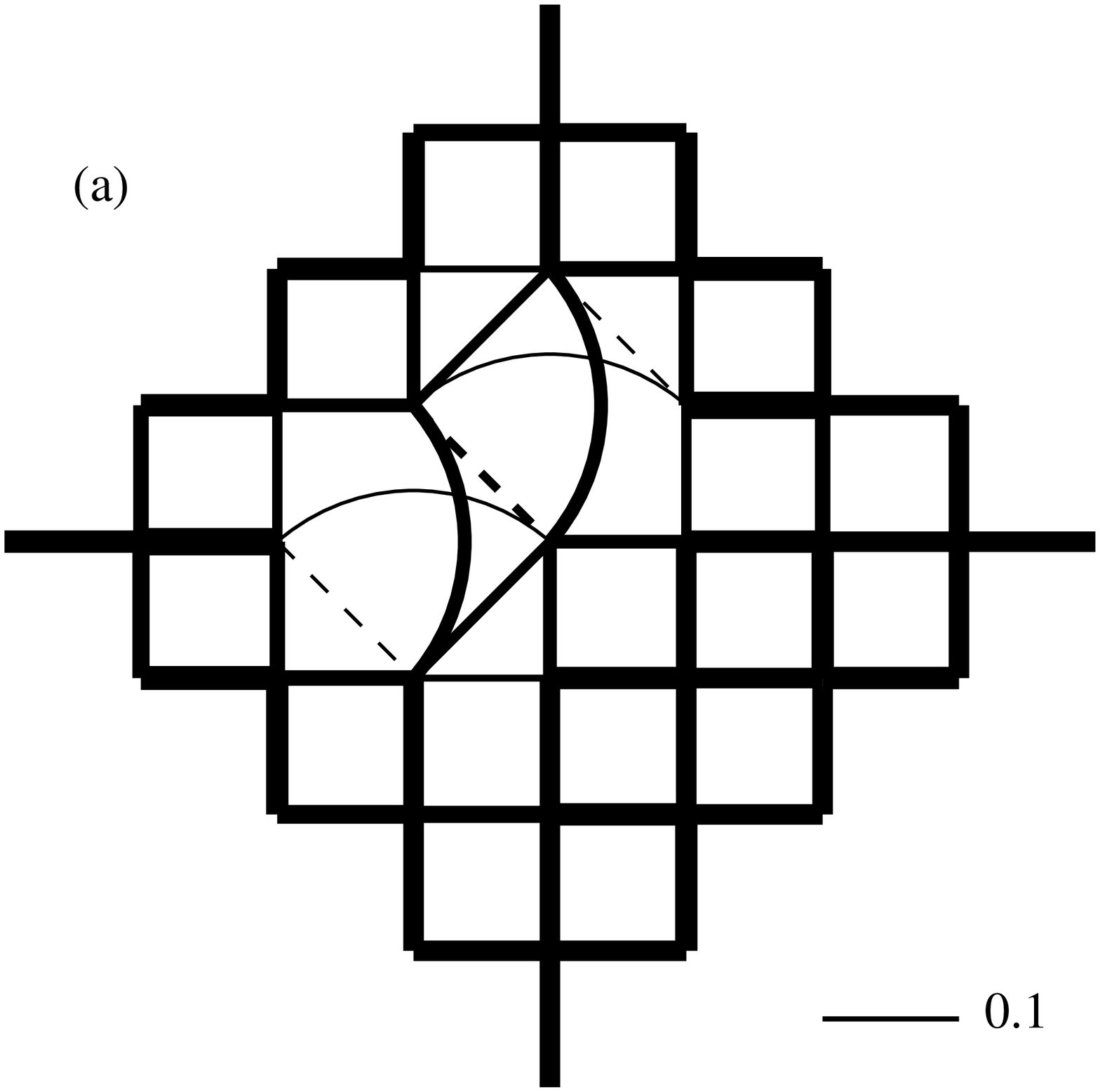,width=4cm}
\psfig{figure=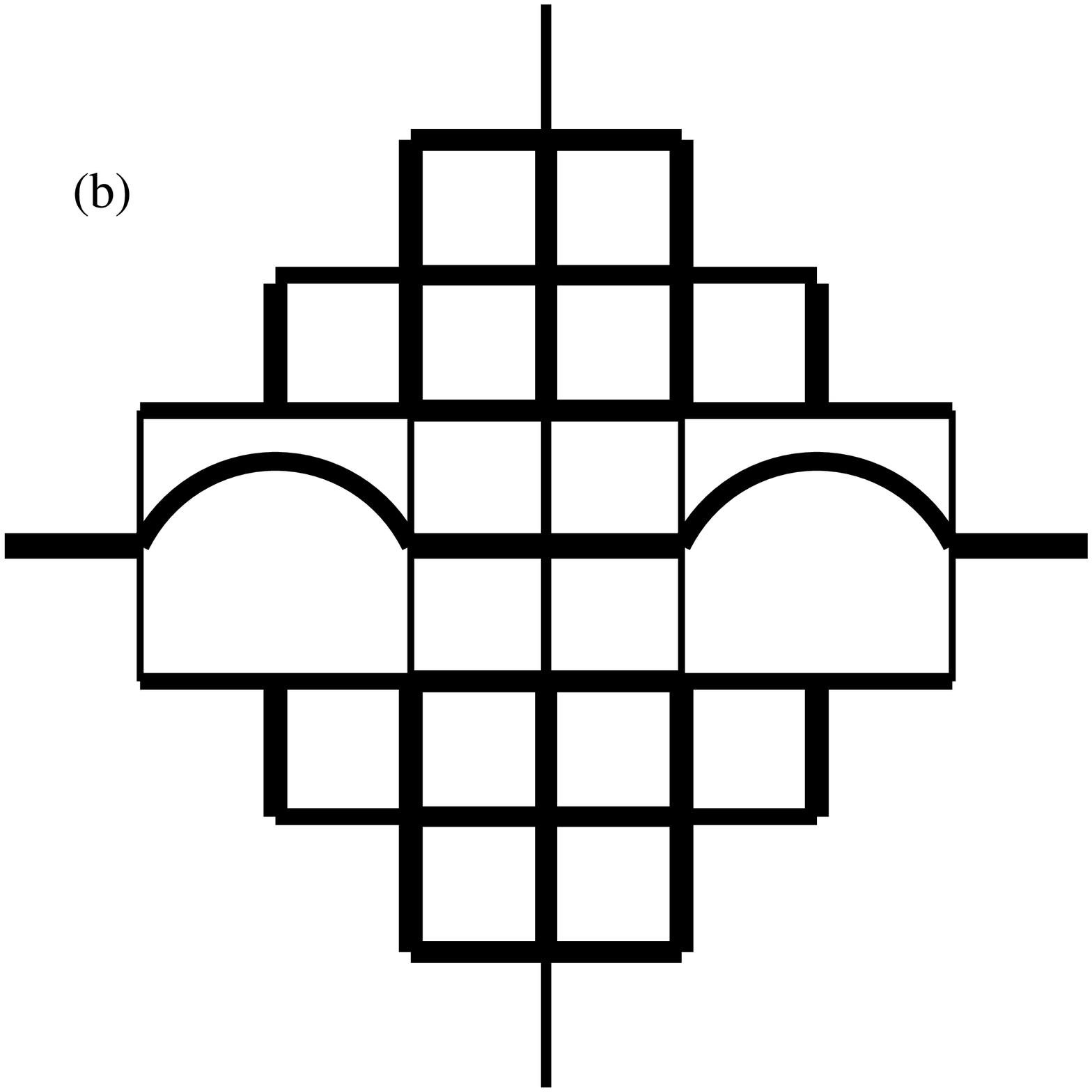,width=4cm}
}
\caption{Same as Fig.~\ref{fig:sisj} but for the $p_{(\pi,\pi)}$ state at
  $J=0.3t$. The scale of the line width is also 
the same as in Fig.~\ref{fig:sisj}.}
\label{fig:sisj2}
\end{figure}

Next let us consider the  low-energy state $p_{(\pi,\pi)}$.
As already pointed out before,\cite{ws97} the spin correlation in the
vicinity of a pair of near-by holes
discussed above provides a mechanism for hole binding. It has also
been argued that such spin correlation is a result of the
$d_{x^2-y^2}$ symmetry of the wavefunction.\cite{rd98}
Then it is reasonable to expect that in wavefunctions with other
symmetries, such mechanism for hole binding is missing and the holes may
be unbound. Fig.~\ref{fig:sisj2} seems to support this.
When we compare the $d$ and $p_{(\pi,\pi)}$ wavefunctions in the same subspace
where the holes are at $\sqrt{2}$ apart (Fig.~\ref{fig:sisj}a and
\ref{fig:sisj2}a), we find striking differences in the spin
correlation. The strong diagonal singlet bond between the holes
found in the $d$ state does not exist in the $p_{(\pi,\pi)}$ state. The binding
mechanism discussed above is therefore missing in the $p_{(\pi,\pi)}$ state.
This causes the
probability for the hole pairs at $\sqrt{2}$ apart to be small, as
evident from Fig.~\ref{fig:hh_corr}. On the other hand, when the holes are
far apart we observe a strong singlet spin pair across each hole
(Fig.~\ref{fig:sisj2}b). It becomes a strong nearest neighbor singlet
bond when the hole hops, thus increasing the overlap with other hole
configurations and making it a preferred hole configuration. This
feature is missing in the $d$ state (Fig.~\ref{fig:sisj}d), where
large hole separations are not preferred. This
explains the repulsive nature of the holes in the $p_{(\pi,\pi)}$ state as shown
in Fig.~\ref{fig:hh_corr}. Note that this interpretation is also
consistent with the hole hopping energy as shown in
Fig.~\ref{fig:ke}b. The hopping energy is larger when the holes are
farther apart.
 
\section{Momentum space distribution}

\begin{figure}
\centerline{
\psfig{figure=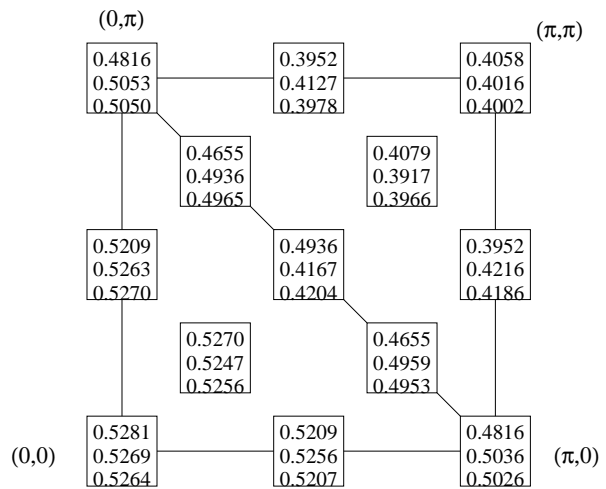,width=8cm}
}
\caption{EMDF of the low energy states.
  Each box represents an allowed momentum {\bf k}. The three
  numbers inside each box, from top to bottom, are $\langle n_{\bf
  k}\rangle$  of
  the $d$, $p_{(\pi,\pi)}$ and $p_{(\pi,0)}$ states at $J=0.3t$. Only the first
  quadrant of the Brillouin zone is shown. $\langle n_{\bf k}\rangle$
  in other quadrants can be constructed by using the
  symmetry of the Brillouin zone.}
\label{fig:nq}
\end{figure}

Fig.~\ref{fig:nq} shows the
 electron momentum distribution function (EMDF), defined as
$\langle n_{{\bf k}\sigma}\rangle\equiv 
\langle \tilde{c}^\dagger_{{\bf k}\sigma}\tilde{c}^{}_{{\bf k}\sigma}\rangle$,
 for the three low
 energy states.
Since in the two-hole model $\langle n_{{\bf
  k}\uparrow}\rangle=\langle n_{{\bf k}\downarrow}\rangle$, 
we will leave out the spin index
$\sigma$. This quantity demonstrates another striking difference between 
the $d$ and $p$
 states. 
Previous works\cite{ew93} have shown that the general shape of
 $\langle n_{\bf k}\rangle$ is irrelevant to the structure of the hole pair.
 This is evident from the fact that in the two-hole model,
 $\Delta n\equiv(\langle n_{(0,0)}\rangle - \langle
 n_{(\pi,\pi)}\rangle)$ is roughly the same as $(\Delta
 n^{1h}_\uparrow+\Delta n^{1h}_\downarrow)$.\cite{1h} Note that
 this is true for all the three low-energy states. 
Nevertheless,  those {\bf k}
 along the magnetic Brillouin zone boundary [$(\pi,0)$ to $(0,\pi)$]
are important in reflecting the
 structure of the bound state.\cite{I}
The deviation of the EMDF from the undoped value, $\langle n_{\bf
 k}\rangle-\frac{1}{2}$, represents the hole weight at that {\bf
 k}. As already pointed out in Ref.~\onlinecite{I} and shown in
 Fig.~\ref{fig:nq}, $\langle n_{\bf k}\rangle$ of the $d$ state along
 the magnetic Brillouin zone boundary has a maximum at $(\pi/2,\pi/2)$, and
 minimum somewhere between $(\pi/2,\pi/2)$ and $(\pi,0)$. Note that in
 the $d$ state $\langle n_{\bf k}\rangle$ is symmetric about the line
 from $(0,0)$ to $(\pi,\pi)$, and in the 32-site lattice the two
 minima
in the first quadrant of the Brillouin zone
 are at $(3\pi/4,\pi/4)$ and $(\pi/4,3\pi/4)$. On the contrary,
 in the $p$ states the minimum of $\langle n_{\bf
 k}\rangle$ along the magnetic Brillouin zone boundary is at
 $(\pi/2,\pi/2)$. Note that the symmetry of the $d$ state requires
  the hole weight at $(\pi/2,\pi/2)$ to be zero. Hence its $\langle
 n_{(\pi/2,\pi/2)}\rangle$ deviates from the undoped value of
 $\frac{1}{2}$ by a minimal value of $0.0064$. Such restriction is
 lifted in the $p$ states and in fact $\langle n_{\bf k}\rangle$ shows
 that their hole weight is a maximum at $(\pi/2,\pi/2)$. This feature
 resembles a hole pocket at $(\pi/2,\pi/2)$.
The fact that the EMDF of the $p$ states have dimples at momenta
 corresponding to the
 single-particle ground state leads us to think that
some form of rigid band filling approximation 
works  in the $p$ states but not in the $d$ state.
This can be accounted for by their
 different  hole-hole correlations in real space. When the holes are
 farther apart, the overlap in the spin distortions they produce is
 smaller. Therefore the state can be better approximated by some
 combination of the single-hole quasiparticle states,
 thus making the rigid band filling model a better approximation.

\section{Spin correlation}

\begin{figure}
\centerline{
\psfig{figure=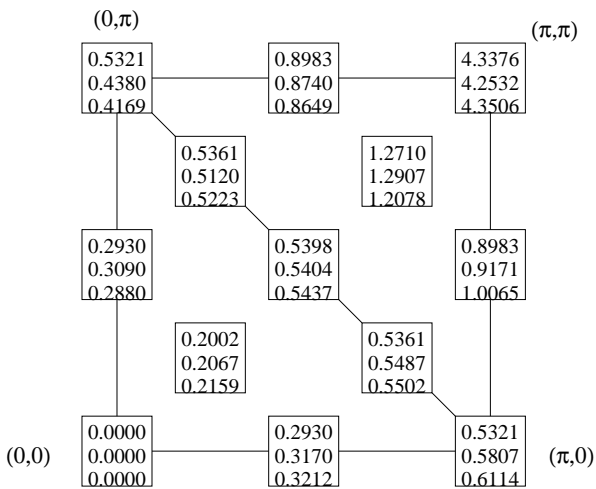,width=8cm}
}
\caption{Same as Fig.~\ref{fig:nq} but for the static structure factor 
$S({\bf k})$.}
\label{fig:sq}
\end{figure}

\begin{figure}
\centerline{
\psfig{figure=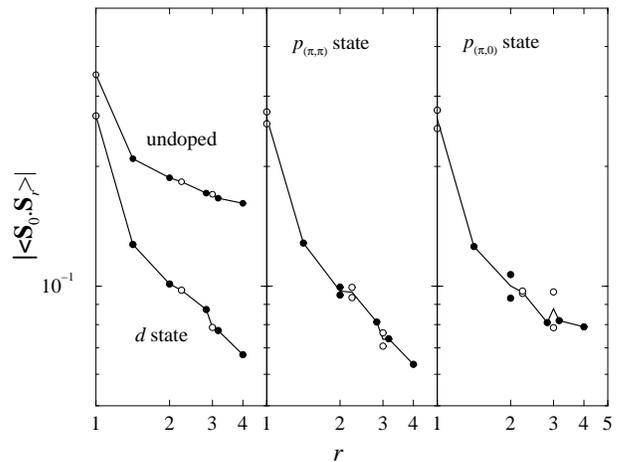,width=8cm}
}
\caption{Spin correlation function $\langle {\bf
  S}_0\cdot{\bf S}_r\rangle$ of the three low energy states at
  $J=0.3t$. Also shown is the same result on the ground state of the
  undoped system on the same lattice. Filled and empty circles
  represent 
  positive and negative
  correlations respectively.}
\label{fig:ss}
\end{figure}

Next we turn to the spin properties of the three low energy states at
$J=0.3t$. We first study the static structure factor $S({\bf k})$ and
the spin correlation function $\langle{\bf S}_0\cdot{\bf
  S}_r\rangle$. They are related by
\begin{equation}
S({\bf k})=\sum_r e^{-i{\bf k}\cdot{\bf r}}\langle{\bf S}_0\cdot{\bf
  S}_r\rangle.
\end{equation}
Fig.~\ref{fig:sq} shows the static structure factors.
All three states show strong characteristics of a N\'eel state --
$S({\bf k})$ strongly peaks at $(\pi,\pi)$. It is interesting
to note that
while those properties associated with the holes are very different
between the $p$ and $d$ states, their static structure factors
are very similar. 
As far as the spin property is concerned, the
 effect of the holes added to the system is to
weaken the N\'eel order only,  with no qualitative change.
This is also indicated in the spin correlation function $\langle {\bf
  S}_0\cdot{\bf S}_r\rangle$  in Fig.~\ref{fig:ss}. It is obvious
that the spin correlations of the $d$ and $p_{(\pi,\pi)}$ states have very
similar behavior. If we fit a power law to the data points at $r\ge
2$, we find that the spin correlation decays as $r^{-0.61}$ in the
$d$ state and $r^{-0.68}$ in the $p_{(\pi,\pi)}$ state. They are to be
compared to the spin correlation of the undoped ground state 
on the same 32-site lattice which
decays as $r^{-0.25}$. The spin correlation
of the $p_{(\pi,0)}$ state seems to be of a bit longer range and cannot be
fitted satisfactorily by a power law. However, the difference does not
show up significantly in the static structure factor.

\begin{figure}
\centerline{
\psfig{figure=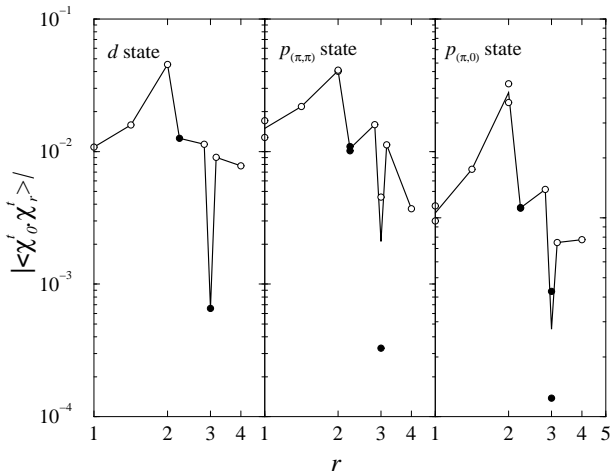,width=8cm}
}
\caption{Same as Fig.~\ref{fig:ss} except for the spiral spin
  correlation $\langle\chi^t_0\cdot\chi^t_r\rangle$.}
\label{fig:twist}
\end{figure}

In the semi-classical theory of Shraiman and Siggia, a mobile hole
produces long-range dipolar distortion in the spin background.\cite{ss88} A
consequence is that the holes tend to stay away to minimize the
overlap of the spin distortion they produce, resulting in dimples in
the EMDF at momenta corresponding to the single-hole ground state. As
discussed in the last paragraph of the previous section, the $p$ states
seem to fit this scenario better than the $d$ state. Another
consequence of the semi-classical theory is that
in the presence of a small number of
holes, the system may be unstable towards the spiral phase.\cite{ss89}
Therefore it is interesting to compare the tendency of the $d$ and $p$
states to have spiral order.
A consequence of the spiral phase is that the maximum of $S({\bf k})$
will shift away from ${\bf k}=(\pi,\pi)$. As shown in
Fig.~\ref{fig:sq}, we do not see such pattern in any of the three
low-energy states. However, the shift in {\bf k}
may be too small to be detected in our finite lattice with discrete
{\bf k}. In previous ED studies of the $t$-$J$ model on smaller
lattices, the shift in $S({\bf k})$ was not observed until the doping
level is much larger.\cite{mdjr90,gvl94} On the other hand,
experiments has demonstrated incommensurability in the magnetic
fluctuation.\cite{incom} Therefore one should not look for long range
order that exists in the thermodynamic limit, but should instead look
for short-range spiral order.
For this reason we calculate the
twist order parameter which is defined as
\begin{equation}
{\bm\chi}^t_i={\bf S}_i\times({\bf S}_{i+\bf x}+{\bf S}_{i+\bf y}).
\label{twist}
\end{equation}
Fig.~\ref{fig:twist} shows 
$\langle{\bm\chi}^t_0\cdot{\bm\chi}^t_r\rangle$ in the
three low energy states. The short range behavior of this 
correlation function is very similar in
the $d$ and $p_{(\pi,\pi)}$ states.
In the $p_{(\pi,0)}$ state it seems to decay
faster. We find no enhancement in the short range spiral order
in the $p$ states relative to the $d$ state,
and the correlation functions are very small in all  three states. The
susceptibility, defined as 
$\langle\frac{1}{N^2}|\sum_r{\bm\chi}^t_r|^2\rangle$, are
$0.02519$, $0.02296$, and $0.02492$ for the $d$, 
$p_{(\pi,\pi)}$ and $p_{(\pi,0)}$
states respectively. These are to be compared to the corresponding
value of the undoped model on the same lattice, $0.02052$.

\section{robustness of the $\lowercase{d_{x^2-y^2}}$ ground state}

Fig.~\ref{fig:delta2} shows that in our system the energies of the $d$ and
 $p$  states are very close at $J\sim0.3t$. This casts doubt on
 whether the $d$ state is always the ground state of the two-hole
 model in the physically relevant range of $J$. Furthermore, the
 small energy difference means that including
additional terms in the $t$-$J$ model Hamiltonian may have profound
effect on the selection of the ground state. This raises the question
 on the robustness of the $d$ state as the ground state.
In this section we discuss the effect
on the energy levels
when two kinds of terms are added to the $t$-$J$ model.

\begin{figure}
\centerline{
\psfig{figure=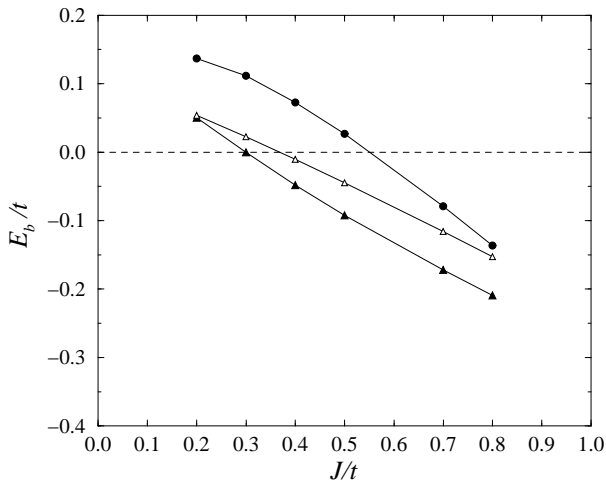,width=8cm}
}
\caption{Same as Fig.~\ref{fig:delta2} except that the Coulomb
  repulsion terms $V_1=0.6t$ and $V_2=0.3t$ are added to the $t$-$J$
  model Hamiltonian. The states shown are $d$ (circles),
$p_{(\pi,\pi)}$ (solid triangles) and $p_{(\pi,0)}$ (empty triangles).}
\label{fig:v1v2}
\end{figure}

As pointed out in Ref.~\onlinecite{rd98}, a small size bound pair may
be destroyed by a realistic short-range Coulomb repulsion. On a
two-leg ladder, the authors have shown that the bound pair disappears
around $V\sim4J$, where $V$ is the nearest neighbor Coulomb repulsion.
Recently Coulomb repulsions are found to enhance the staggered orbital
current on a two-leg ladder.\cite{fm01} On a two-dimensional lattice,
such staggered current has been found even without  Coulomb
repulsions.\cite{ilw00,l00} 
Adding them to the two-dimensional model may do more than just pushing
the holes apart.
The repulsions should weaken  hole binding and
raise the energy of the $d$ state. But in
the $p$ states, the effect is less significant because the holes are already
far apart. As a result one would expect  short-range Coulomb
repulsions to favor the $p$ states to the $d$ state. 
Here we add two repulsion terms $V_1$
and $V_2$ to the $t$-$J$ Hamiltonian. They are repulsions between two
holes  at 1 and $\sqrt{2}$ apart respectively. To test
the stability of the $d_{x^2-y^2}$ ground state, we choose to use the
 values $V_1=0.6t$, $V_2=0.3t$. The binding energies of the  three
states are shown in Fig.~\ref{fig:v1v2} as a function of $J/t$. 
As expected, the binding
energy of the $d$ state is pushed up more than that of the $p$ states. 
At $J=0.3t$, only $E_b$ of the $p_{(\pi,\pi)}$ state is
barely negative. In the range we study, $0.2t\le J\le0.8$, the $p_{(\pi,\pi)}$
state has the lowest energy.
This calculation shows that even a small
Coulomb repulsion will destroy the two-hole bound state and select the
$p$ state as the ground state.

It is know that in order to reproduce the single hole dispersion
observed experimentally,\cite{wells} one has to add  longer range
hole hopping terms
$t'$ and $t''$ to the $t$-$J$ model.\cite{nvhdg95,lwg97} These terms
are relevant here for two reasons. First the magnitude of these
terms, $t'=-0.3t$ and $t''=0.2t$,\cite{lwg97} are larger than the energy
difference between the $d$ and $p$ states
at $J=0.3t$. Therefore they may have
serious effect on the low-lying energy levels. Second, it has been argued
that the form of single-hole dispersion resulting from the $t'$ and
$t''$ terms favors the $p$ state.\cite{cl99} In
Ref.~\onlinecite{lwg97},  three-site hopping terms are also
included in addition to $t'$ and $t''$
in order to reproduce the spectral
functions measured experimentally. However, if we just want to
reproduce the single hole dispersion, it will be suffice to include 
 the $t'$ and $t''$
terms only.  
Table~\ref{tab:tp} shows the lowest energy state in
different symmetry subspaces when $t'=-0.3t$ and $t''=0.2t$ are
included in the $t$-$J$ model at $J=0.3t$.
\begin{table}
\caption{Energies $E_{2h}$ and binding energies $E_b$  
of the lowest energy states with different symmetries of
  the two-hole $t$-$t'$-$t''$-$J$ model with $J=0.3t$, $t'=-0.3t$, and
  $t''=0.2t$. Ground state energy $E_{1h}$ of the one-hole 
model is $-13.716526t$.}
\label{tab:tp}
\begin{ruledtabular}
\begin{tabular}{cr@{\,}lcdd}
symmetry&\multicolumn{2}{c}{momentum}&spin&\multicolumn{1}{c}{$E_{2h}/t$}&\multicolumn{1}{c}{$E_b/t$}\\
\colrule
$d$       &(0,&0)               &0 &-15.588993   &0.51434\\
$s$       &(0,&0)               &0 &-16.074058   &0.02927\\
$p$       &(0,&0)               &1 &-16.080823   &0.02251\\
$p$       &($\pi$,0),&$(0,\pi)$ &0 &-16.109653   &-0.00632\\
$p$       &($\pi$,&$\pi$)       &0 &-16.164444   &-0.06111\\
\end{tabular}
\end{ruledtabular}
\end{table}
We can see that the structure of the low-lying energy levels is
seriously affected. 
First, the energy of the
$d$ state is pushed up by a lot compared to  other states.
Its energy is now even higher than an $s$ and a (triplet) $p$ state.
Note that  in the $t$-$J$ model these two states have higher energies
than the $d$ state (see section \ref{sec:energy}).
Nevertheless, their  binding energies are positive. 
Only the two singlet $p$ states have negative binding
energies. The $p$ state with momentum $(\pi,\pi)$ now becomes the
ground state. Note that its binding energy does not differ too much
from that of the $t$-$J$ model as shown in Fig.~\ref{fig:delta2},
which is $-0.05146t$.

\section{Conclusion}

We tried to map out the low-lying states of the $t$-$J$ model with two
holes on a 32-site square lattice at various $J/t$. While previous studies have
focussed on finding the critical $J/t$ where hole binding is lost, we
concentrate on the symmetry of the ground state at different $J/t$. We
find that at large $J/t$, the ground state has $d_{x^2-y^2}$ symmetry
as reported before. But there are low-lying states with $p$ symmetry
whose energies come closer to the ground state as $J/t$ decreases. At
$J/t<0.3$ one of these $p$-wave state becomes the ground state. We
note that these $p$-wave states have non-zero momenta. Previous
numerical works have revealed similar finite momentum states. But calculations
carried out on  lattices smaller than 32 sites are biased by their
geometry. As a result, these
finite momentum $p$ states have not received much
attention. Nevertheless, a similar cross-over in the symmetry of the
ground state from $d$ to $p$ symmetry has been reported in a series
expansion study of the $t$-$J$ model.\cite{hwo98} 
There the cross-over was found to be at
$J/t\sim0.4$, which is quite close to our value of $J/t\sim0.3$. This
cross-over was also observed in the anisotropic $t$-$J_z$
model.\cite{hwo98,cl99} 
Although finite-size effects prevent us from concluding definitely
whether this cross-over exists in the thermodynamic limit, the fact
that it occurs within the parameter range of interest make us feel that the
$p$ states are relevant to the low energy physics of the $t$-$J$ model. 
Further investigation reveals qualitative differences and similarities
in the properties of the $p$ and $d$ states.
Perhaps the most intriguing is the difference in their hole properties.
The holes are mutually attractive in the $d$ state, but are
repulsive in the $p$ states. The spin structure in the vicinity of the
holes provides an intuitive explanation of the contrasting
behaviors. And as a result of this, the electron momentum distribution
functions are also very different. While the EMDF of the $d$ state 
shows no sign of hole pockets, that of the $p$ states clearly have
dimples at $(\pi/2,\pi/2)$. 
Such hole properties  suggest that the rigid-band filling
assumption should work better in the $p$ than in the $d$ state, if it
works at all. And in the framework of the semi-classical
theory,\cite{ss89} one would expect the $p$ states to have stronger tendency
to show  spiral spin order than  the $d$ state.
However, we do not find
any enhancement in the short-range spiral spin correlation in the $p$
states over the $d$ state. In fact
the spin correlations of the $d$ and $p$ states are very similar and
show signals of the N\'eel state.

We have also demonstrated that the $d_{x^2-y^2}$ ground state is
not robust. It can be destroyed easily by including realistic terms to
the $t$-$J$ model Hamiltonian. Both the short-range Coulomb repulsion 
and  longer range hopping terms ($t'$ and $t''$)
favor the $p$ state with momentum $(\pi,\pi)$ as the ground
state. This shows that hole pairing in the $d_{x^2-y^2}$ channel may
not be a generic feature of the $t$-$J$ model. There is a competing $p$
state which has no hole pairing.
The implication of our result is that the symmetry of the two-hole
ground state of the $t$-$J$ model is yet to be determined.
This is in contrary to a recent numerical study\cite{sorella}
 which concludes that
pairing in the $d_{x^2-y^2}$ channel is a robust property of the $t$-$J$ model.

\begin{acknowledgements}

We thank A. L. Chernyshev for very helpful discussions.
This work was supported by the RGC
of Hong Kong under Grant No. HKUST6146/99P. 
\end{acknowledgements}


\begin{thebibliography}{}
\bibitem{pstate} To be exact, the $p_{(\pi,\pi)}$ state has $p_x$ symmetry and
  is degenerate with another state with $p_y$ symmetry and the same
  momentum. The $p_{(\pi,0)}$ state also has $p_x$ symmetry, and is degenerate
  with another state with $p_y$ symmetry and momentum $(0,\pi)$.
\bibitem{prob} The probability $P(r)$ of finding the hole at $r$ apart
  is related to the hole-hole correlation function by
  $P(r)=N_E(r)C(r)$, where $N_E(r)$ is the number of equivalent pairs
  at the same $r$ apart in this finite cluster.
\end{thebibliography}
\end{document}